\documentclass[twocolumn,showpacs,amsmath,amssymb,pra]{revtex4}
\usepackage{graphicx}
\usepackage{dcolumn} 
\usepackage{bm}      

\begin{document}

\title{Completely Positive Bloch-Boltzmann Equations}
\author{Robert Alicki}           \email{fizra@univ.gda.pl}
\author{Stanis{\l}aw Kryszewski} \email{fizsk@univ.gda.pl}
\affiliation{Institute of Theoretical Physics and Astrophysics,
             University of Gda{\'n}sk, ul. Wita Stwosza 57,
             80-952 Gda{\'n}sk, Poland}
\date{\today}
\begin{abstract}
The density operator of the arbitrary physical system must be
positive definite. Employing the general master equation technique
which preserves this property we derive equations of motion for the
density operator of an active atom which interacts collisionally with
the reservoir of perturber atoms. The obtained general relations
applied to the two-level atom yield Bloch-Boltzmann equations (BBE).
The form of the BBE obtained by us differs from that known from
literature which, as we show, are not guaranteed to preserve the
required positivity. We argue that our results are the correct ones
and as such should be used in practical applications. Moreover, the
structure and the terms which appear in our set of ~BBE~ seem to
allow simpler and more straightforward physical interpretation.
\end{abstract}
\maketitle


\newcommand{\vv}{\wek{v}}
\newcommand{\ckk}[1]{{\cal K}_{#1}(\vv \leftarrow \vv')}
\newcommand{\wkk}[1]{{\cal W}_{#1}(\vv \leftarrow \vv')}
\newcommand{\half}{{\textstyle \frac{1}{2}}}
\newcommand{\ddt}{\frac{d}{dt}\:}
\newcommand{\op}[1]{\hat{#1}}
\newcommand{\ops}[1]{\hat{#1}^{\dagger}}
\newcommand{\leftz}{\left( \rule[0.0mm]{0mm}{3.4mm}}
\newcommand{\leftk}{\left[ \rule[0.0mm]{0mm}{3.4mm}}
\newcommand{\wek}[1]{\vec{\bf {#1}}}
\newcommand{\bra}[1]{\langle \, #1 \, |} 
\newcommand{\ket}[1]{| \, #1 \, \rangle} 
\newcommand{\elm}[3]{\bra{#1} \, #2 \, \ket{#3}}

\section{Introduction}

A lot of experiments  in  atomic  physics  and  spectroscopy consists
in investigating active atoms coupled to incident (laser) radiation
and immersed in the thermal bath of perturbers which, typically, are
the atoms of noble gas. Various phenomena occurring in such a system
as well as its properties are then investigated. The amount of work
devoted to such studies is enormous, it is therefore quite impossible
even to list all relevant literature, except a few essential
monographs reviewing the subject \cite{rash,scul}.

The theoretical description of the discussed system must account for
two major aspects. Firstly, the coupling of active atoms to the
incoming radiation field and the radiative spontaneous phenomena must
be properly described. This is usually done by means of standard
methods of quantum optics \cite{cohen,puri}. As a result one  obtains
a set of equations of motion for the matrix elements of the atomic
density operator. When the atom is described within a two-level model
the obtained equations are known as optical Bloch equations.
Secondly, active atoms undergo collisions both with perturbers and
among themselves. The influence of collisions on physical properties
of the system constituents is of paramount importance and is in
itself a separate field of experimental and theoretical studies.
Discussion of these problems in their full generality clearly goes
beyond the scope of the present work. Let us however mention, that in
spectroscopical applications the effect of collisions is usually
accounted for by suitably derived (quantum-mechanically or
classically) collision terms. Historically speaking, Boltzmann was
the first to introduce the collision terms into the equations of
motion of the probability distributions. Therefore, for a two-level
atom, the combination of optical Bloch equations together with
collision terms might be called Bloch-Boltzmann equations (BBE) which
account for both kinds of the discussed interactions influencing the
behavior of active atoms. However, we feel it necessary to stress
that the name {\em Bloch-Boltzmann equations} need not be restricted
to two-level atoms. Generalizations to more complex atomic models do
not pose serious conceptual difficulties, though the form of
corresponding equations of motion might be much more complicated.

The main aim of this paper is to reexamine the origin and form of the
collision terms in Bloch-Boltzmann equations. We shall mostly study a
simple two-level model, because it is formally the simplest, it
allows the simplest interpretations, yet retaining the most important
(at least in a qualitative manner) features of realistic physical
situations and experiments. The~motivation for our research is the
following. First of all we note, that the density operator of an
arbitrary physical system must always be positive definite. It is not
clear whether the collision terms, used within the literature which
is known to us, have this property. Moreover, they seem to exhibit
other drawbacks or inconsistencies. These terms are derived
(quantum-mechanically or classically) by using arguments similar to
those leading to collision integrals of standard Boltzmann equation
\cite{rash}. We shall try to present a consistent theory which will,
hopefully, allow us to clarify the question of positive definiteness
as well as some more subtle points.

The tools necessary to construct the proper form of BBE describing
the system (active atoms) coupled to a reservoir (perturbers) are
provided by the quantum theory of dynamical semigroups which entail
the general master equation (ME) methods. It is worth stressing that
we have in mind mathematically rigorous version of the ME theory
based on completely positive quantum dynamical semigroups.
The extensive review of this subject is given in  monographs
\cite{alic,alfa}, where the authors derive and discuss the most
general (sometimes called the Lindblad form \cite{gks,lind}) ME which
preserves the positivity of the considered density operator. Sec. II
will, therefore, be devoted to a brief review of the essentials of
this theory. In Sec. III we discuss how the general theory can be
formally adapted to describe a mixture of two gaseous species when
various models of internal structure of active atoms can be employed.
We outline the procedures necessary to find explicit expressions for
formal quantities introduced in the preceding section. We also try to
identify physical assumptions needed to validate the presented
theory. We restrict our attention to two-level atoms and, hence, in
Sec. IV we give the derivation of the formal Bloch-Boltzmann
equations in the above discussed sense. Finally, Sec.V is devoted to
the discussion of our results in the view of the facts known from the
literature. Some features of the obtained BBE are different from
those known from other sources. We argue that our results are the
proper ones. Since we employ much different theoretical formalism,
the comparison of our results with other ones seems to be of an
essential importance.

As we have already stated, we focus our attention on the collisional
phenomena occurring in the active-atom-perturber gaseous mixture.
Therefore, we leave the radiative effects out of the picture. We
shall, however, briefly indicate how, and under what conditions, such
effects can be incorporated back into our formalism.

\section{Theoretical framework}

The physical system we will consider in this work, is a mixture of
active atoms $(A)$ with density $N_{A}$ and perturbers of density
$N_{p}$. We assume that $N_{A} \ll N_{p}$, which allows us to neglect
the $A-A$ collisions. Thus, only $A-P$ collisions can affect the
motion of the active atoms. Moreover, we assume that the density of
perturbers is such, that only binary $A-P$ collisions are of
importance. The relatively dense perturber gas acts as a reservoir of
energy and momentum and influences the velocity distributions of the
active atoms. On the other hand, it is usually safe to assume that
$P-P$ collisions are frequent enough to assure rapid thermalization
of the perturbers. This allows us to think of a perturber bath as
being in thermal equilibrium, so that the velocity distribution of
$P$ atoms is time-independent and given by a Maxwellian
\begin{equation}
  W_{p}(\wek{v}) =  \left( \frac{1}{\pi u_{p}^{2}} \right) ^{3/2}
          \exp \left( - \frac{ \wek{v}^{2}}{u_{p}^{2}} \right),
\label{ppmax} \end{equation}
with $u_{p}^2 = 2k_{B} T/m_{p}$ being the square of the most probable
velocity of perturber atoms with mass $m_{p}$, at temperature $T$.
The stated physical conditions are not really very restrictive. They
are fairly well satisfied by a great number of realistic experiments
\cite{rash}.

\subsection{Master equation for the quantum-classical system}

The master equation techniques stem from the theory of quantum
dynamical semigroups applied to an open system, that is to a system
which interacts with another one which serves as a reservoir. The
active atoms may be considered as the quantum-mechanical open system
which is coupled to the reservoir consisting of the perturbers. The
interaction between the two subsystems is manifested by the
collisions occurring between $A$ and $P$ particles. The proper
theoretical framework for the description of an open system, which
ensures the preservation of positivity of the reduced density
operator of the system of interest (active atoms), is supplied by the
master equation (ME) approach \cite{alic}. It is not our aim to
review the theory or derivation of ME. We shall rather apply the
general ME to a class of open systems which can be called
quantum-classical ones. We study a quantum-mechanical open system,
the states of which span the Hilbert space
\begin{equation}
    {\cal H} = \bigoplus_{\alpha} {\cal H}_{\alpha}.
\label{hsp1} \end{equation}
The index $\alpha$ belongs to a set ${\cal A}$ which, for current
purposes, is assumed to be discrete, but will subsequently be
generalized to continuous one. A more physical interpretation of the
spaces ${\cal H}_{\alpha}$ will be given later. We assume that within
space ${\cal H}$ there exists a strong decoherence mechanism
\cite{blan,petr,joos} which practically excludes quantum
superpositions of the form
\begin{equation}
   \ket{\psi_{\alpha_{1}}}
       \oplus \ket{\psi_{\alpha_{2}}}
       \oplus \ldots \ldots ~.
\label{excl} \end{equation}
This is due to environmental decoherence mechanism, which is a
generally accepted explanation of the absence of superpositions of
macroscopically distinguishable states (Schr{\"o}dinger cat problem)
and emergence of classical properties, called dynamical
superselection rules. The most effective physical mechanism leading
to strong decoherence is provided by many subsequent collisions of
the particles of the investigated system with the environment
particles \cite{alic}.

On each space $ {\cal H}_{\alpha}$ we define a reduced density
operator $\rho_{\alpha}$. The family of such operators
\begin{equation}
   \rho = \{ \rho_{\alpha} \}_{\alpha \in {\cal A}},
\label{rodef} \end{equation}
forms partially diagonal, quantum-classical density operator which
describes the properties of the relevant (open) system. The operators
$\rho_{\alpha}$ have the following properties
\begin{eqnarray}
    \mathrm{(i)}
    && \rho_{\alpha} : {\cal H}_{\alpha}
       \rightarrow {\cal H}_{\alpha};
\nonumber \\
    \mathrm{(ii)}
    && \rho_{\alpha} \ge 0,
    \hspace*{5mm} \mathrm{positive-definiteness};
\nonumber \\
    \mathrm{(iii)}
    && \sum_{\alpha} \mathrm{Tr} \rho_{\alpha} = 1,
    \hspace*{5mm} \mathrm{normalization}.
\label{rocon} \end{eqnarray}
\hspace*{7mm}The most general form of the Markovian master equation
for our quantum-classical density operator can be obtained from the
general form of the generator of the completely positive quantum
dynamical semigroup (Lindblad-Gorini-Kossakowski-Sudarshan form). The
relevance of complete positivity in the theory of quantum open
systems is extensively discussed in Refs.\cite{alic,alfa}. The
corresponding master equation derived along these lines, which
governs the evolution of the quantum-classical density operator of
the relevant system due to its interaction with the reservoir, reads
(in the Schr{\"o}dinger picture)
\begin{eqnarray}
   \ddt \rho_{\alpha}
   = &-& \frac{i}{\hbar} \leftk \op{H}_{\alpha},\rho_{\alpha} \right]
       +  \sum_{\beta} \sum_{\xi}
              \op{S}_{\alpha \beta}^{\,\xi} \rho_{\beta}
             (\op{S}_{\beta \alpha}^{\,\xi} )^{\dagger}
\nonumber \\
     &-& \half \leftz \op{B}_{\alpha}\rho_{\alpha}
         + \rho_{\alpha} \op{B}_{\alpha} \right),
\label{megen} \end{eqnarray}
where additional index $\xi$ allows full flexibility to describe
various dissipation phenomena. The operators introduced in this
equation are defined as mappings:
\begin{subequations} \label{menot}
\begin{eqnarray}
    \mathrm{(i)}
    && \op{H}_{\alpha} = \ops{H}_{\alpha}
       : {\cal H}_{\alpha} \rightarrow {\cal H}_{\alpha},
       ~~\mathrm{(hamiltonian)};
\label{menota} \\
    \mathrm{(ii)}
    && \op{S}_{\alpha \beta}^{\,\xi}
       : {\cal H}_{\beta} \rightarrow {\cal H}_{\alpha};
\label{menotb} \\
    \mathrm{(iii)}
    && (\op{S}_{\beta \alpha}^{\,\xi})^{\dagger}
       : {\cal H}_{\alpha} \rightarrow {\cal H}_{\beta},
       ~~\mathrm{a ~map ~dual ~to} ~\op{S}_{\alpha \beta}^{\,\xi};
\label{menotc} \\
    \mathrm{(iv)}
    && \op{B}_{\alpha}
    = \sum_{\xi} \sum_{\beta}
      (\op{S}_{\alpha \beta}^{\,\xi})^{\dagger}
      \op{S}_{\beta \alpha}^{\,\xi}.
\label{menotd}
\end{eqnarray} \end{subequations}
The specific form of the Hamiltonian $\op{H}_{\alpha}$ depends on the
particular physical properties of the studied relevant system.
Operators $\op{S}_{\alpha \beta}^{\,\xi}$ depend on the interaction
between the relevant subsystem and the reservoir. Recently the
equation of this type has found its application in the quantum
measurement theory (see the contribution of Blanchard and Jadczyk in
\cite{blan}, where its properties are also  widely discussed).

\subsection{Master equation for an atom immersed in perturber gas}

General and rather formal master equation (\ref{megen}) has to be
adapted to describe the presently discussed system, that is the
moving (with velocity $\wek{v}$) atom which collides with the
perturbers. We shall proceed taking care of any additional or
simplifying assumptions which go beyond the ones adopted in the
derivation of ME (\ref{megen}). Moreover, we will proceed in a
manner, which may be useful when considering active atoms with the
internal structure more general than the simple two-level model.

The collisions with perturber particles lead to strong decoherence
which together with the uniform spatial distribution of interacting
particles justify the use of the density operators which are diagonal
in momentum (or velocity) representation. Therefore, the discrete
decomposition in Eq.(\ref{hsp1}) can be replaced by a continuous one
which is taken to be with respect to the velocity $\wek{v}$ of an
active atom. Thus, Eq.(\ref{hsp1}) is modified and becomes
\begin{equation}
   {\cal H} = \int d\vv ~{\cal H}_{\vv},
   \hspace*{5mm}\mathrm{with}\hspace*{5mm}
   \vv \in \mathbb{R}^{3}.
\label{hspv} \end{equation}
The ensemble of active particles is now described by a partially
diagonal density operator $\rho(\wek{v})$. We associate the space
${\cal H}_{\vv}$ with the state space of an active atom which
possesses velocity $\wek{v}$. We introduce a set of operators
$\{S_{a}\}$, which constitutes a basis in the space of relevant
operators acting on ${\cal{H}}_{\vv}$. The specific form of operator
basis depends on the model chosen to describe the internal structure
of an active atom. One may choose a multilevel model for which one
has $S_{a}=S_{kl}=\ket{k}\bra{l}$, with $k,l=1,2,\ldots,n$, and with
$\ket{k}$ being the energy eigenstates. Alternatively, spherical
tensor operators might be taken as a basis which is appropriate for
atoms with spatially degenerate energy levels. Later on, we will
consider a simple two-level model and we will explicitly define the
necessary operator basis.

First we analyze the hamiltonian term in (\ref{megen}). To this end
we expand it in the operator basis $\{S_{a}\}$, and we write
\begin{equation}
   \frac{1}{\hbar} H_{\alpha} ~~\longrightarrow~~
   \frac{1}{\hbar} H(\vv) = \sum_{a} h_{a}(\vv) S_{a}.
\label{hf1} \end{equation}
The particular form of the functions $h_{a}(\vv)$ need not be
specified now. By the proper choice of these functions we can model
various physical situations, some of which will be discussed later.
For now, the first (hamiltonian) term of master equation becomes
\begin{equation}
   - i \sum_{a=1}^{4} h_{a}(\vv)
        \leftk S_{a}, ~\rho(\vv) \right].
\label{me1a} \end{equation}
It must be, however, noted that the Hamiltonian $H(\vv)$ should be
hermitian, so the functions $h_{a}(\vv)$ must satisfy some additional
conditions, the particular form of which depend on the choice of the
operator basis. We shall illustrate this point when applying the
general formalism to derivation of the Bloch-Boltzmann equations for
a two-level atom.

Since the indices $\alpha$ and $\beta$ are replaced by the
"classical" degrees of freedom, that is by velocities, when
constructing the second term of ME (\ref{megen}) we must replace the
summation over the index $\beta$ by integration over velocities.
Following the general rules given in \cite{alic} we may rewrite the
second term in the ME as
\begin{eqnarray}
   &&\sum_{\beta} \sum_{\xi}
        \op{S}_{\alpha \beta}^{\,\xi} \rho_{\beta}
       (\op{S}_{\beta \alpha}^{\,\xi})^{\dagger}
\nonumber \\
   &&~~\longrightarrow~~
   \sum_{a,b} \int d\vv'
       ~\ckk{ab} S_{a} \rho(\vv') S_{b}^{\dagger}.
\label{me2a} \end{eqnarray}
This term obviously has the sense of an operator which describes the
transitions from a velocity group around $\vv'$ to the velocity
interval $(\vv, \vv + d\vv)$. Hence it can be called a "gain" term.
We shall postpone the discussion of the integral kernel to the
further sections. At present, according to relation (\ref{me2a}),
we shall only require that for any velocities $\wek{v}$
and $\wek{v}'$
\begin{equation}
   \ckk{ab} ~-~ \mathrm{positively ~defined ~matrix},
\label{kpdef} \end{equation}
of the necessary dimensions. This matrix contains the details of the
collisional interaction between the active and perturber atoms which
will also be discussed later.

Following further the principles of the construction of the ME
\cite{alic}, we proceed to the third term in (\ref{megen}). It is an
anticommutator and it is built similarly to the former one. Namely,
it can be rewritten as
\begin{eqnarray}
  && \hspace*{-7mm}
   - \half \leftz \op{B}_{\alpha}\rho_{\alpha}
         + \rho_{\alpha} \op{B}_{\alpha} \right)
\nonumber \\
  && \hspace*{-7mm}
     \longrightarrow~
     - \half \sum_{a,b} \int d\vv'
        ~{\cal K}_{ab}^{\ast} (\vv' \leftarrow \vv)
         \{ S_{a}^{\dagger} S_{b}, ~\rho(\vv) \},
\label{me3} \end{eqnarray}
where the curly brackets denote an anticommutator. This term also
describes the transition -- escape from a velocity group $(\vv, \vv +
d\vv)$ to any other velocity, so it is a "loss" term.

Combining the discussed three terms, we now construct the master
equation for a density operator of the moving active atom. We note,
that no additional approximations (apart from those involved in the
derivation of the general ME (\ref{megen})) were made. Thus we have
\begin{eqnarray}
   && \hspace*{-1mm}
      \ddt \rho(\vv)
      = - i \sum_{a} h_{a}(\vv) \leftk S_{a}, ~\rho(\vv) \right]
\nonumber \\
   && \hspace*{20mm}
    + \sum_{a,b} \int d\vv' ~\ckk{ab} S_{a} \rho(\vv') S_{b}^{\dagger}
\nonumber \\
   && \hspace*{20mm}
    - \half \sum_{a,b} \int d\vv'
       ~{\cal K}_{ab}^{\ast} (\vv' \leftarrow \vv)
         \{ S_{a}^{\dagger} S_{b}, ~\rho(\vv) \}.
\label{meful} \end{eqnarray}
Let us note that the integration over velocity in the last term,
affects only the integral kernel. Hence we can introduce the rate
\begin{equation}
    \gamma^{\ast}_{ab} \equiv  \gamma^{\ast}_{ab}(\vv)
    = \int d\vv'
       ~{\cal K}_{ab}^{\ast} (\vv' \leftarrow \vv),
\label{kgam} \end{equation}
with the aid of which, our ME becomes
\begin{eqnarray}
   \ddt \rho(\vv)
   =&-&i\sum_{a} h_{a}(\vv) \leftk S_{a}, ~\rho(\vv) \right]
\nonumber \\
   &-&\half \sum_{a,b}  \gamma^{\ast}_{ab}(\vv)
           \{ S_{a}^{\dagger} S_{b}, ~\rho(\vv) \}
\nonumber \\
  &+&\sum_{a,b} \int d\vv' ~\ckk{ab} S_{a} \rho(\vv') S_{b}^{\dagger}.
\label{meff} \end{eqnarray}
The obtained ME is an operator equation. For practical purposes it
is then convenient to expand the density operator $\rho(\vv)$ in
suitably selected operator basis. One can then compute all the
necessary operator commutators and products, thus obtaining the
equations of motion for the matrix elements of the density operator.
We shall do so in the further section, by adopting a simple two-level
model. Nevertheless, we stress that the presented ME (\ref{meff}) can
be employed for atomic models more general than just the simple
two-level one.

\section{Microscopic derivation of ~$\ckk{ab}$}

The formalism so far presented is fairly general. We proceed with its
further discussion and clarification. The gain term (\ref{me2a}) and
the loss one (\ref{me3}) which describe the irreversible evolution of
the relevant system stem from its interaction with environment. In
our case, collisions are the manifestation of this interaction. The
physical details concerning collisions are hidden in, so far rather
formal, collision kernels $\ckk{ab}$ which were left unspecified.
Certainly, their structure and mathematical properties follow from
the procedures used when deriving the necessary master equation.

Derivation of the master equation for an open quantum system from the
underlying fundamental Hamiltonian dynamics was  the subject of very
many investigations. Although the number of relevant literature
sources is enormous, in only a few of them proper care is taken with
respect to mathematical consistency of the presented results. The
density operator of an arbitrary system (interacting with the
surroundings, or not) should be positive definite. This can be
ensured only by the carefully taken and properly conducted limiting
procedures. It is not our aim to review these rigorous mathematical
techniques such as weak coupling (or van Hove method) \cite{davies},
singular coupling \cite{goko} or low density limit \cite{dum},
we refer the reader to \cite{alic} for a survey of the subject.

We shall briefly discuss only the last of the mentioned limiting
procedures -- the low density limit which is designed specially for
the description of a quantum system interacting collisionally with a
low density perturber gas. The underlying physical assumptions are
that only binary collisions are of importance and that the duration
of the collision is much shorter than the mean free-flight time (this
latter condition is the essence of the so-called impact
approximation). The reasoning leading to specific form of the
collision kernels is thus, as follows. Let us temporarily assume that
the considered active atom is confined within a finite volume $V$
and, therefore, is described by discrete states $\ket{\vv, j}$ with
discrete velocities and with quantum number $j$ denoting its internal
state (note that $j$ may serve as a multiindex consisting of several
quantum numbers). These states span certain Hilbert space, which by
an introduction of an energy cut-off can be made finite
dimensional. Let us say that $N$ is the (finite) number of the basis
vectors $\ket{\vv, j}$. Then, we can use the completely positive
generator obtained for a $N$-state quantum system interacting with
dilute gas of perturbers in the low density limit \cite{dum}.
Finally, one takes the density operator which is partially diagonal
in velocities, removes the energy cut-off and takes the limit $V
\longrightarrow \infty$ (a kind of a thermodynamic limit). Thus one
arrives at the master equation which formally coincides with
Eq.(\ref{meff}). The discussed procedure allows us to assign concrete
meaning to all the terms which appear in the master equation.

The Hamiltonian of the active atom is of the form
\begin{equation}
   H^{(0)} = \sum_{j}^{n} E_{j} \; \ket{j}\bra{j}
           = \sum_{j}^{n} \hbar \omega_{j} \; \ket{j}\bra{j},
\label{haa} \end{equation}
where some eigenenergies $E_{j} = \hbar \omega_{j}$ may be degenerate
and where $n$ fixes the dimension of space of internal states of the
active atom. This restricts the most general form of the Hamiltonian
(\ref{hf1}) to a diagonal one. Operators $S_{a}$ employed earlier,
may be specified as eigensolutions to the equation
\begin{equation}
   \leftk H^{(0)}, ~S_{a} \right] = \hbar \Omega_{a} S_{a},
   \hspace*{10mm} a = 1,2, \ldots\ldots,n^{2},
\label{sdef} \end{equation}
with $\Omega_{a}$ denoting the differences between atomic
eigenfrequencies. Finally, the low density limit in the interaction
terms leads to the following expression for the collision kernels
\begin{widetext}
\begin{eqnarray}
   \ckk{ab}
   &=& \frac{(2\pi)^{4}\:\hbar^{2}}{\mu^{3}}
       ~N_{p} \; \delta_{\Omega_{a},\Omega_{b}}
       \int d\vv'_{r} \int d\vv_{r}
      ~\delta^3 \left[ \vv - \vv'
        -\frac{\mu}{m} \left(\vv_{r} - \vv'_{r} \right) \right]
\nonumber \\
&& \times
    ~\delta \left( \frac{\mu v_{r}^{2}}{2}
         - \frac{\mu {v_{r}'}^{2}}{2} + \hbar \Omega_{a} \right)
    ~W_{p}(\vv' - \vv_{r}')
    ~{T}_{a}( \vv_{r} \leftarrow \vv_{r}') \:
     {T}_{b}^{\ast}(\vv_{r} \leftarrow \vv_{r}'),
\label{ker1} \end{eqnarray}
\end{widetext}
where $\vv$ (or $\vv'$) are the velocities of an active atom, and
$\vv_{p}$ (or $\vv_{p}'$) of the perturber  after (or before)
collision. $\vv_{r} = \vv - \vv_{p}$ (or $\vv'_{r} = \vv'-\vv_{p}'$)
are the corresponding relative velocities. $m$, $m_{p}$ and $\mu$
denote the masses of the active atom, perturber and the reduced mass,
respectively. $W_{p}(\vv_{p}')$ describes the equilibrium velocity
distribution of the perturber atoms, so in most cases it is just a
Maxwellian as given in Eq.(\ref{ppmax}). The functions
${T}_{a}(\vv_{r} \leftarrow \vv_{r}')$ are related to the standard
$\op{T}$-matrix (known from the quantum-mechanical scattering theory)
in the center of mass variables
\begin{equation}
   \sum_{a} \; T_{a}(\vv_{r} \leftarrow \vv_{r}') S_{a}
   = \sum_{i,j=1}^{n} \elm{\vv_{r},j}{\op{T}}{\vv_{r}',j'}
      ~\ket{j}\bra{j'}.
\label{tdef} \end{equation}
The resulting matrix $\ckk{ab}$ is block-diagonal due to degeneracies
$\Omega_{a} = \Omega_{b}$ which is ensured by the Kronecker delta
$\delta_{\Omega_{a},\Omega_{b}}$ and obviously positively definite.

For purposes of the further discussion it is convenient to express
matrix $\ckk{ab}$ by standard scattering amplitudes. From quantum
scattering theory \cite{taylor}, we recall that the scattering
amplitudes and functions $T_{a}(\vv_{r} \leftarrow \vv_{r}')$ are
connected by the relation
\begin{equation}
   f_{a}(\vv_{r} \leftarrow \vv_{r}')
   = - (2\pi)^{2} \: \frac{\hbar}{\mu^{2}}
      ~T_{a}(\vv_{r} \leftarrow \vv_{r}').
\label{scam} \end{equation}
Hence, rewriting Eq.(\ref{ker1}) we alternatively have
\begin{widetext}
\begin{eqnarray}
   \ckk{ab}
   &=& \mu N_{p} \; \delta_{\Omega_{a},\Omega_{b}}
       \int d\vv'_{r} \int d\vv_{r}
      ~\delta^3 \left[ \vv - \vv'
        -\frac{\mu}{m} \left(\vv_{r} - \vv'_{r} \right) \right]
\nonumber \\
&& \hspace*{3mm} \times
    ~\delta \left( \frac{\mu v_{r}^{2}}{2}
         - \frac{\mu {v_{r}'}^{2}}{2} + \hbar \Omega_{a} \right)
    ~W_{p}(\vv' - \vv_{r}')
    ~f_{a}( \vv_{r} \leftarrow \vv_{r}') \:
     f_{b}^{\ast}(\vv_{r} \leftarrow \vv_{r}').
\label{ker2} \end{eqnarray}
\end{widetext}
Moreover, it might be also convenient to note that the delta function
reflecting energy conservation, can be written in several forms,
each of them being used by different authors. These forms are
\begin{eqnarray}
   && \hspace*{-10mm}
     \delta \left( \frac{\mu v_{r}^{2}}{2}
                - \frac{\mu {v_{r}'}^{2}}{2} + \hbar \Omega_{a} \right) =
\nonumber \\
  && \hspace*{10mm}
    = \frac{2}{\mu}
        ~\delta \left( v_{r}^{2} - {v_{r}'}^{2}
              + \frac{2 \hbar \Omega_{a}}{\mu} \right)
\nonumber \\
  && \hspace*{10mm}
    = \frac{1}{\mu v_{r}}
       ~\delta \left( \sqrt{ v_{r}^{2}  + \frac{2 \hbar \Omega_{a}}{\mu} }
           - v_{r}' \right).
\label{vardel} \end{eqnarray}

\hspace*{7mm} Equation (\ref{meff}), together with the Hamiltonian
specified in Eq.(\ref{haa}) and with collision kernels given by
(\ref{ker1}) or (\ref{ker2}) govern the evolution of the density
operator of active atom interacting collisionally with perturbers. It
preserves automatically positivity and normalization of the partially
diagonal density operator $\rho(\wek{v})$.

\section{Bloch-Boltzmann equations}

\subsection{Introductory remarks}

Our considerations presented in previous sections can be applied to
multilevel atoms with complex internal structure. This is done by a
suitable definition of operators $S_{a}$ introduced in
Eq.(\ref{sdef}). Such a choice also specifies the form of the
collision kernels. We shall, however, illustrate the general approach
by its application to a two-level atom which is a typical model for
many quantum-optical phenomena. The resulting equations are, as we
already mentioned, called Bloch-Boltzmann equations. In this case,
the space ${\cal{H}}_{\vv}$ of an atom having velocity $\wek{v}$ is
isomorphic with $\mathbb{C}^{2}$. We denote the ground state by
$\ket{1}$ and the excited state by $\ket{2}$. Taking them to form a
basis in ${\cal H}_{\vv}$, we adopt the following identifications
\begin{equation}
     \ket{1} ~=~
         \left( \begin{array}{c} 0 \\ 1 \end{array} \right),
\hspace*{20mm}
     \ket{2} ~=~
         \left( \begin{array}{c} 1 \\ 0 \end{array} \right).
\label{vec} \end{equation}
The space of operators acting in ${\cal H}_{\vv}$ is thus spanned by
$2 \times 2$ matrices. As a basis in the operator space we choose
following four operators (pseudospin matrices)
\begin{subequations} \label{psd}
\begin{eqnarray}
   \ket{1} \bra{1}
   &=& \left( \begin{array}{cc} 0 & 0 \\ 0 & 1 \end{array} \right)
   \equiv S_{1} = S_{1}^{\dagger},
\label{psda} \\
   \ket{2} \bra{2}
   &=& \left( \begin{array}{cc} 1 & 0 \\ 0 & 0 \end{array} \right)
   \equiv S_{2} = S_{2}^{\dagger},
\label{psdb} \\
   \ket{2} \bra{1}
   &=& \left( \begin{array}{cc} 0 & 1 \\ 0 & 0 \end{array} \right)
   = S_{+} \equiv S_{3} = S_{4}^{\dagger},
\label{psdc} \\
   \ket{1} \bra{2}
   &=& \left( \begin{array}{cc} 0 & 0 \\ 1 & 0 \end{array} \right)
   = S_{-} \equiv S_{4} = S_{3}^{\dagger}.
\label{psdd}
\end{eqnarray} \end{subequations}
The right-hand sides of these equations also fix the notation we will
use in this paper. For sake of easy reference we quote the nonvanishing
commutators of operators $S_{a}$:
\begin{equation}
\renewcommand{\arraycolsep}{5mm}
\begin{array}{ll}
     \leftk \: S_{1}, ~S_{3} \right]= -S_{3},
   & \leftk \: S_{1}, ~S_{4} \right]= S_{4},  \\[2mm]
     \leftk \: S_{2}, ~S_{3} \right]= S_{3},
   & \leftk \: S_{2}, ~S_{4} \right]= -S_{4},  \\[2mm]
   & \leftk \: S_{3}, ~S_{4} \right]= S_{2}-S_{1}.
\end{array}
\renewcommand{\arraycolsep}{1mm}
\label{pscom} \end{equation}
Any operator on $~{\cal H}_{\vv}$ can be expressed by the basis ones,
while its matrix elements would be parameterized by velocity $\vv$.
In particular, for a general Hamiltonian introduced in (\ref{hf1}),
we have
\begin{eqnarray}
   \frac{1}{\hbar} H(\vv)
   &=& h_{1}(\vv) S_{1} + h_{2}(\vv) S_{2}
     + h_{3}(\vv) S_{3} + h_{4}(\vv) S_{4}
\nonumber \\
   &=& \renewcommand{\arraycolsep}{3mm}
       \left( \begin{array}{cc}
         h_{1}(\vv) & h_{4}(\vv) \\[2mm]
         h_{3}(\vv) & h_{2}(\vv)
       \end{array} \right),
\renewcommand{\arraycolsep}{1mm}
\label{htla} \end{eqnarray}
what is in accord with notation introduced in (\ref{psd}). We note,
that the Hamiltonian must be Hermitian, which implies that functions
$h_{1}(\wek{v})$, ~$h_{2}(\wek{v})$ are real, while
$h_{3}(\wek{v}) = h_{4}^{\ast}(\wek{v})$.
Density operator $\rho(\vv)$ can be expanded exactly in the same
manner. To fix the notation and terminology we, therefore, write
\begin{eqnarray}
   \rho(\vv)
   &=& \rho_{11}(\vv) \ket{1} \bra{1}
     + \rho_{12}(\vv) \ket{1} \bra{2}
\nonumber \\
  && \hspace*{5mm}+ \rho_{21}(\vv) \ket{2} \bra{1}
     + \rho_{22}(\vv) \ket{2} \bra{2}
\nonumber \\
   &=& \rho_{1}(\vv) S_{1}
     + \rho_{2}(\vv) S_{2}
\nonumber \\
  && \hspace*{5mm}+ \rho_{3}(\vv) S_{3}
     + \rho_{4}(\vv) S_{4}.
\label{roexp} \end{eqnarray}
Hence, we can identify the matrix elements
\begin{subequations} \label{roel}
\begin{eqnarray}
    \rho_{1}(\vv) = \rho_{11}(\vv),
    && \hspace*{4mm}
    \rho_{2}(\vv) = \rho_{22}(\vv),
\label{roela} \\
    \rho_{3}(\vv) = \rho_{21}(\vv),
    && \hspace*{4mm}
    \rho_{4}(\vv) = \rho_{12}(\vv),
\label{roelb}
\end{eqnarray} \end{subequations}
where the first row gives the populations, while the second one --
coherences, which specify the physical meaning of the matrix elements
of the considered density operator.

Finally, we note that relation (\ref{haa}) restricts the general form
of the hamiltonian (\ref{htla}) to $h_{3}=h_{4}=0$, that is to
\begin{equation}
\renewcommand{\arraycolsep}{3mm}
   \frac{1}{\hbar} H^{(0)}
   = \left( \begin{array}{cc}
          \omega_{1} &     0       \\[2mm]
              0      & \omega_{2}  \end{array} \right),
\renewcommand{\arraycolsep}{1mm}
\label{htla2} \end{equation}
where $E_{j}=\hbar \omega_{j}$, ~$(j=1,2)$, are the eigenenergies of
the corresponding levels of the active atom. Furthermore, it is
straightforward to see that operators $S_{a}$ defined in (\ref{psd})
satisfy relations (\ref{sdef}) with hamiltonian (\ref{htla2}).
The eigenvalue $\Omega_{1}=\Omega_{2}=0$ is doubly degenerate, while
$\Omega_{3}=\omega_{21}$, and $\Omega_{4}=-\omega_{21}$ (where
$\hbar\omega_{21}=\hbar(\omega_{2}-\omega_{1}) > 0$, is the energy
difference between two atomic levels).

\subsection{General form of Bloch-Boltzmann equations}

Next step of our derivation consists in the expansion of the density
operator of a two-level atom according to Eq.(\ref{roexp}). Doing so
in the both sides of Eq.(\ref{meff}) we then perform all the
necessary operator computations. As a result we arrive at the
following set of equations for each matrix element (\ref{roel}) of
the density operator
\begin{widetext}
\begin{subequations} \label{bbe}
\begin{eqnarray}
   \ddt \rho_{1}(\vv)
   &=& - i h_{4}(\vv) \rho_{3}(\vv)
       + i h_{3}(\vv) \rho_{4}(\vv)
\nonumber \\
   && - \leftz \gamma^{\ast}_{11}
              + \gamma^{\ast}_{33} \right) \rho_{1}(\vv)
      - \half \leftz \gamma^{\ast}_{41}
              + \gamma^{\ast}_{23} \right) \rho_{3}(\vv)
      - \half \leftz \gamma^{\ast}_{32}
              + \gamma^{\ast}_{14} \right) \rho_{4}(\vv)
\nonumber \\
   && + \int d\vv' \leftk
        \ckk{11} \rho_{1}(\vv') + \ckk{44} \rho_{2}(\vv')  \right.
\nonumber \\
   && \hspace*{30mm} \left.
      + \ckk{41} \rho_{3}(\vv') + \ckk{14} \rho_{4}(\vv')
        \rule[0.0mm]{0mm}{3.4mm} \right],
\label{bbea} \\[3mm]
   \ddt \rho_{2}(\vv)
   &=&   i h_{4}(\vv) \rho_{3}(\vv)
       - i h_{3}(\vv) \rho_{4}(\vv)
\nonumber \\
    && - \leftz \gamma^{\ast}_{22}
              + \gamma^{\ast}_{44} \right) \rho_{2}(\vv)
       - \half \leftz \gamma^{\ast}_{41}
              + \gamma^{\ast}_{23} \right) \rho_{3}(\vv)
       - \half \leftz \gamma^{\ast}_{32}
              + \gamma^{\ast}_{14} \right) \rho_{4}(\vv)
\nonumber \\
   && + \int d\vv' \leftk
        \ckk{33} \rho_{1}(\vv')  + \ckk{22} \rho_{2}(\vv') \right.
\nonumber \\
   && \hspace*{30mm} \left.
      + \ckk{23} \rho_{3}(\vv') + \ckk{32} \rho_{4}(\vv')
        \rule[0.0mm]{0mm}{3.4mm} \right],
\label{bbeb} \\[3mm]
   \ddt \rho_{3}(\vv)
   &=& - i h_{3}(\vv) \leftz \rho_{1}(\vv) - \rho_{2}(\vv) \right)
       + i \leftz h_{1}(\vv) - h_{2}(\vv) \right) \rho_{3}(\vv)
\nonumber \\
   &&  - \half \leftz \gamma^{\ast}_{32}
              + \gamma^{\ast}_{14} \right) \rho_{1}(\vv)
       - \half \leftz \gamma^{\ast}_{32}
              + \gamma^{\ast}_{14} \right) \rho_{2}(\vv)
\nonumber \\
   && \hspace*{30mm}
       - \half \leftz \gamma^{\ast}_{11}
                    + \gamma^{\ast}_{22}
                    + \gamma^{\ast}_{33}
                    + \gamma^{\ast}_{44} \right) \rho_{3}(\vv)
\nonumber \\
   && + \int d\vv' \leftk
        \ckk{31} \rho_{1}(\vv') + \ckk{24} \rho_{2}(\vv') \right.
\nonumber \\
   && \hspace*{30mm} \left.
      + \ckk{21} \rho_{3}(\vv') + \ckk{34} \rho_{4}(\vv')
        \rule[0.0mm]{0mm}{3.4mm} \right],
\label{bbec} \\[3mm]
   \ddt \rho_{4}(\vv)
   &=&   i h_{4}(\vv) \leftz \rho_{1}(\vv) - \rho_{2}(\vv) \right)
       - i \leftz h_{1}(\vv) - h_{2}(\vv) \right) \rho_{4}(\vv)
\nonumber \\
   && - \half \leftz \gamma^{\ast}_{23}
              + \gamma^{\ast}_{41} \right) \rho_{1}(\vv)
       - \half \leftz \gamma^{\ast}_{23}
              + \gamma^{\ast}_{41} \right) \rho_{2}(\vv)
\nonumber \\
   && \hspace*{30mm}
       - \half \leftz \gamma^{\ast}_{11}
                    + \gamma^{\ast}_{22}
                    + \gamma^{\ast}_{33}
                    + \gamma^{\ast}_{44} \right) \rho_{4}(\vv)
\nonumber \\
   && + \int d\vv' \leftk
        \ckk{13} \rho_{1}(\vv') + \ckk{42} \rho_{2}(\vv') \right.
\nonumber \\
   && \hspace*{30mm} \left.
        + \ckk{43} \rho_{3}(\vv') + \ckk{12} \rho_{4}(\vv')
       \rule[0.0mm]{0mm}{3.4mm} \right]
\label{bbed}.
\end{eqnarray} \end{subequations}
\end{widetext}
This set of equations is the most general one, describing the
evolution of active atoms (within a two-level model) due to the
interaction with environment -- collisions with the perturber atoms.
Since these equations are derived directly from the general ME which
preserves the positivity of the reduced density operator of $A$
atoms, we may be certain that this property is unchanged.

The obtained BBE are somewhat simpler if we take into account the
hamiltonian (\ref{htla2}). The first lines of Eqs.(\ref{bbea}) and
(\ref{bbeb}) disappear, while the first lines of Eqs.(\ref{bbec}) and
(\ref{bbed}) become $-i\omega_{21} \rho_{3}(\wek{v})$ and
$i\omega_{21} \rho_{4}(\wek{v})$, respectively. In the next
subsections we will briefly present some frequently used
approximations which allow some simplifications of the general form
(\ref{bbe}) with the first lines modified, as it was just discussed.
We will first consider the so-called secular approximation and
afterwards we will assume that inelastic collisions can be neglected.
However, it seems that the sequence of these approximations is
irrelevant and that each of them can be used independently of the
other one.

\subsection{Bloch-Boltzmann equations and secular approximation}

Secular approximation is recognized as a useful tool in the analysis
of master equation technique applied to a manifold of physical
systems. Its validity in quantum-optical problems is thoroughly
discussed by Cohen-Tannoudji \cite{cohen} and by Puri \cite{puri}.
The essence of this approximation can, in the present context, be
summarized as follows. When the off-diagonal terms in the Hamiltonian
(\ref{htla}) are much smaller (or simply negligible) as compared to
the diagonal ones, the populations and the coherences may be
approximately decoupled. In the other words, under the secular
approximation, the populations are coupled only to populations, while
each of the coherences is coupled only to itself. So far, we have
considered active atoms which are not subjected to any other
interactions apart from the collisions. Since we have taken
the hamiltonian (\ref{htla2}), the secular approximation is
clearly justified, which entails considerable simplification of the
general Bloch-Boltzmann equations. Thus, we obtain
\begin{widetext}
\begin{subequations} \label{bbs}
\begin{eqnarray}
   \ddt \rho_{1}(\vv)
   &=& - \leftz \gamma^{\ast}_{11}(\vv)
              + \gamma^{\ast}_{33}(\vv) \right) \rho_{1}(\vv)
       + \int d\vv' \leftk \ckk{11} \rho_{1}(\vv')
             + \ckk{44} \rho_{2}(\vv') \right],
\label{bbsa} \\[3mm]
   \ddt \rho_{2}(\vv)
   &=& - \leftz \gamma^{\ast}_{22}(\vv)
              + \gamma^{\ast}_{44}(\vv) \right) \rho_{2}(\vv)
       + \int d\vv' \leftk
         {\cal K}_{33}(\vv \leftarrow \vv') \rho_{1}(\vv')
       + {\cal K}_{22}(\vv \leftarrow \vv') \rho_{2}(\vv') \right],
\label{bbsb} \\[3mm]
   \ddt \rho_{3}(\vv)
   &=& - i \omega_{21} \rho_{3}(\vv)
       - \half \leftz \gamma^{\ast}_{11}(\vv)
                    + \gamma^{\ast}_{22}(\vv)
                    + \gamma^{\ast}_{33}(\vv)
                    + \gamma^{\ast}_{44}(\vv) \right) \rho_{3}(\vv)
       + \int d\vv' ~\ckk{21} \rho_{3}(\vv'),
\label{bbsc} \\[3mm]
   \ddt \rho_{4}(\vv)
   &=& i \omega_{21} \rho_{4}(\vv)
       - \half \leftz \gamma^{\ast}_{11}(\vv)
                    + \gamma^{\ast}_{22}(\vv)
                    + \gamma^{\ast}_{33}(\vv)
                    + \gamma^{\ast}_{44}(\vv) \right) \rho_{4}(\vv)
       + \int d\vv' ~\ckk{12} \rho_{4}(\vv').
\label{bbsd}
\end{eqnarray} \end{subequations}
\end{widetext}
It should be noted, that if the active atoms are also irradiated by
incident radiation the Hamiltonian (\ref{htla2}) will be modified by
the suitable coupling terms. Moreover, one has also to account for
spontaneous emission, i.e., for the coupling to the vacuum field.
This would lead to the appearance of additional terms describing
radiative effects. In such a case the validity of the secular
approximation must be separately investigated.

\subsection{No inelastic collisions}

The energy transfer during the typical collision between the atoms in
the gaseous mixture is of the order of $k_{B}T$. Since the typical
temperatures of spectroscopical experiments are of the order of
several hundred kelvins, the energy available during the collision is
by two orders of magnitude smaller than the typical separation of the
levels of the active atom. Hence, the probability that the collision
would excite or deexcite the active atom is negligible. This means
that the inelastic collisions can be left out of the picture.
Inspection of Eqs.(\ref{bbe}) or (\ref{bbs}) tells us that relations
\begin{equation}
    \mathrm{for} ~a=3,4 ~:~
    {\cal K}_{aa}(\vv \leftarrow \vv') = 0,
    \hspace*{5mm} \Longrightarrow
    \hspace*{5mm} \gamma^{\ast}_{aa} =0,
\label{incol} \end{equation}
are equivalent to neglecting the inelastic collisions. Employing
such an assumption in Eqs.(\ref{bbs}) we get
\begin{widetext}
\begin{subequations} \label{bbn}
\begin{eqnarray}
   \ddt \rho_{1}(\vv)
   &=& - \gamma^{\ast}_{11}(\vv) \rho_{1}(\vv)
       + \int d\vv' ~\ckk{11} \rho_{1}(\vv'),
\label{bbna} \\[2mm]
   \ddt \rho_{2}(\vv)
   &=& - \gamma^{\ast}_{22}(\vv) \rho_{2}(\vv)
       + \int d\vv' ~\ckk{22} \rho_{2}(\vv'),
\label{bbnb} \\[2mm]
   \ddt \rho_{3}(\vv)
   &=& - i \omega_{21} \rho_{3}(\vv)
       - \half \leftz \gamma^{\ast}_{11}(\vv)
               + \gamma^{\ast}_{22}(\vv) \right) \rho_{3}(\vv)
       + \int d\vv' ~\ckk{21} \rho_{3}(\vv'),
\label{bbnc} \\[2mm]
   \ddt \rho_{4}(\vv)
   &=& + i \omega_{21} \rho_{4}(\vv)
       - \half \leftz \gamma^{\ast}_{11}(\vv)
               + \gamma^{\ast}_{22}(\vv)  \right) \rho_{4}(\vv)
       + \int d\vv' ~\ckk{12} \rho_{4}(\vv').
\label{bbnd}
\end{eqnarray} \end{subequations}
\end{widetext}
Bloch-Boltzmann equations  in this form seem to be used in
quantum-optical applications which are known to us. Our derivation
clarifies the procedures leading to these equations. It also
specifies the methods allowing explicit computation both of collision
kernels and rates. Moreover, general form (\ref{bbe}) of the BBE
ensures preservation of the positivity of the density operator of
active atoms due to the conditions (\ref{kpdef}) and (\ref{kgam})
which seem not to be used in the literature.

\section{Discussion}

\subsection{Review of the results known from literature}

The theoretical models employed to describe atomic $A-P$ collisions
which tend to thermalize the velocity states of active atoms are
usually based on the suitable adaptation of the quantum-mechanical,
or classical Boltzmann equation. The most extensive and thorough
review of such an approach is given in a monograph by Rautian and
Shalagin \cite{rash}. These authors present quite general formalism
which can be applied to describe various physical situations
concerning active atoms with complex, multilevel structure of
degenerate energy eigenstates. Moreover, general expressions can also
treat both elastic and inelastic collisions. Since we are mainly
concerned with a simple two-level atom suffering elastic collisions,
we will restrict our attention only to such a model. The derivation
of the collisional terms, as discussed in \cite{rash} is performed by
means of a truncation of the BBGKY hierarchy of equations of motion
to one- and two-particle density operators. Assuming that the active
atoms and perturbers are uncorrelated for time $t \rightarrow -
\infty$, and treating their translational motion quasi classically,
one then arrives at the kinetic equation which governs the evolution
of the matrix elements of the density operator due to collisions.
This equation is of the general form
\begin{eqnarray}
    \left. \ddt \rho_{ij}(\vv) \right|_{coll}
    = &-& \gamma_{ij}(\vv) \: \rho_{ij}(\vv)
\nonumber \\
      &+& \int d\vv' ~\wkk{ij} \: \rho_{ij}(\vv'),
\label{linbol} \end{eqnarray}
for $i,j=1,2$. It is valid for binary elastic collisions in the
impact approximation and for perturbers with homogeneous spatial
distribution. The collisional kernel $\wkk{ij}$ following from
quantum-mechanical Boltzmann equation is expressed via the
corresponding scattering amplitudes (see \cite{rash}, Eq.(2.147))
\begin{widetext}
\begin{eqnarray}
   \wkk{ij}
   & = & 2 \: N_{p} \int d\vv'_{r} \int d\vv_{r}
        ~\delta^3 \left[ \vv - \vv'
             -\frac{\mu}{m} \left(\vv_{r} - \vv'_{r} \right) \right]
\nonumber \\[2mm]
&& \times ~\delta \left( v_{r}^{2} - {v_{r}'}^{2} \right)
    ~W_{p}(\vv' - \vv'_{r})
     ~f_{i}( \vv_{r} \leftarrow \vv'_{r}) \:
      f_{j}^{\ast}(\vv_{r} \leftarrow \vv'_{r}).
\label{qmker} \end{eqnarray}
\end{widetext}
The notation employed here is exactly the same as that in
Eq.(\ref{ker1}), while $f_{i}(\vv_{r} \leftarrow \vv'_{r})$ is the
elastic scattering amplitude of an atom in the internal state
$\ket{i}$. The collision rate $\gamma_{ij}(\vv)$, known from
literature \cite{rash,hube}, has the form
\begin{eqnarray}
   \gamma_{ij}(\vv)
   = && N_{p} \: \frac{2 \pi \hbar}{i \mu} \int d\vv_{r} \:
        W_{p}(\vv - \vv_{r})
\nonumber \\
     && \times ~\left[ f_{i}( \vv_{r} \leftarrow \vv_{r})  -
              f_{j}^{\ast}( \vv_{r} \leftarrow \vv_{r}) \right],
\label{qmrate} \end{eqnarray}
so it is given by the difference between the scattering amplitudes
corresponding to forward scattering.

The diagonal kernels $\wkk{ii}$ and rates $\gamma_{ii}$ are seen to
be real and can be thought of to possess classical limits.
Eq.(\ref{linbol}) multiplied by $d\vv$ (for $i=j$) represents the
collisional diffusion out of and the flow into velocity interval
$(\vv, \vv + d\vv)$ due to scattering of active atoms in $\ket{i}$
state. Therefore, integral term in Eq.(\ref{linbol}) is the gain one
and it gives the number of active atoms in state $\ket{i}$ which
change velocity from $\vv'$ before, to $\vv$ after the collision.
Hence, the collision kernels $\wkk{ii}$ are measures of transition
probabilities from $\vv'$ to $\vv$ velocity groups. The rates
$\gamma_{ii}(\vv)$ enter the loss terms, and give the number of atoms
in the state $\ket{i}$ escaping from velocity interval
$(\vv,\vv+d\vv)$ to any other one. Let us note that
$\gamma_{ii}(\vv)$ can be also viewed as the collision frequency and
its inverse $\tau_{ii}(\vv) = 1/\gamma_{ii}(\vv)$ can be interpreted
as the average time between collisions. Finally, it is worth noting
that the given probabilistic interpretation of the kernel and
frequency is fully consistent with the relation
\begin{equation}
    \gamma_{ii}(\vv) = \int d\vv_{1} ~\wkk{ii},
\label{gcon} \end{equation}
which reflects the requirement of the particle number conservation
and follows from Eqs.(\ref{qmker}) and (\ref{qmrate}) by the
application of the optical theorem.

The quantum-mechanical (linear) Boltzmann equation (\ref{linbol})
specifies the time evolution not only of the diagonal elements of the
density operator, but also of the coherences -- the off-diagonal
ones, that is for $i \neq j$. In such a case, neither the kernels
$\wkk{ij}$ nor the rates $\gamma_{ij}$ have probabilistic
interpretation. The kinetic equation (\ref{linbol}) for coherences
can be recast into the following form
\begin{eqnarray}
    \left. \ddt \rho_{ij}(\vv) \right|_{coll}
    = &-& \gamma_{ij}^{(ph)}(\vv) \: \rho_{ij}(\vv)
      - \gamma_{ij}^{(vc)}(\vv) \: \rho_{ij}(\vv)
\nonumber \\
      &+&\int d\vv' ~\wkk{ij} \: \rho_{ij}(\vv').
\label{cohkin} \end{eqnarray}
The collisional rate is split into two terms:
$\gamma_{ij}^{(ph)}(\vv)$ being called the "phase-changing" rate,
while $\gamma_{ij}^{(vc)}(\vv)$ is the "velocity changing" one
\cite{hube}. The former rate, $\gamma_{ij}^{(ph)}(\vv)$, is the one
usually associated with the homogeneous linewidth which appears in
the typical pressure broadening theories of the atomic spectral
lineshape. It can be argued \cite{hube} that, in many practical
applications, $\gamma_{ij}^{(ph)}$ may be treated as velocity
independent (although this dependence entails interesting
correlations between Doppler and power broadening). Two last terms in
(\ref{cohkin}) describe velocity changing collisions and are
frequently omitted from the equations of motion of density operator.
There are some arguments \cite{hube} justifying such an omission.
However, these arguments are somewhat vague and heuristic, and
according to our knowledge, there is no clear-cut and generally
accepted criterion which specifies when one is allowed to drop out
the velocity changing terms in (\ref{cohkin}).

In many theoretical approaches used to describe the spectroscopical
experiments the semiclassical model is thought to be sufficient. In
such a case, the motion of atoms is taken to be purely classical,
while the internal degrees of freedom are treated
quantum-mechanically. Then, the classical kinetic theory is quite
appropriate for determination of velocity distributions of not too
dense gases. The discussed gas is then considered as a mixture of the
following species: perturbers, active atoms in the ground state and
atoms in the excited state. Since the perturbers are assumed to be in
thermal equilibrium, one can invoke linear Boltzmann equation
\cite{cerci} and derive kinetic equation for classical velocity
distributions $F_{j}(\wek{v})$ which are (up to a normalization
constant) proportional to the probability of an active atom being in
state $\ket{j}$ and possessing velocity $\wek{v}$, that is, to
populations $\rho_{ii}(\vv)$. The resulting kinetic equation has the
form identical to diagonal equations (\ref{linbol}). The classical
kernel for an active atom in state $\ket{j}$ is then
\begin{eqnarray}
   {\cal W}_{j}^{(cl)} (\vv \leftarrow \vv')
     & = &  N_{p}  \int d\vv_{r} \int d\vv_{r}'
     ~\frac{d\sigma_{j}(v_{r},\vartheta)}{d\Omega(\vartheta)}
     ~W_{p}(\vv_{p}')
\nonumber \\[2mm]
&& \hspace*{-15mm}
    \times ~\frac{\delta(v_{r} - v_{r}')}{v_{r}}
    ~\delta^{3} \! \left[ \vv - \vv'
      - \frac{\mu}{m} \left( \vv_{r} - \vv_{r}' \right) \right],
\label{clker} \end{eqnarray}
and the corresponding classical collisional rate is
\begin{equation}
    \gamma_{j}^{(cl)}(\vv)
    = N_{p} \int d\vv_{p} \int d\Omega(\vartheta)
      ~|\vv_{r}| ~W_{p}(\vv_{p})
      ~\frac{d \sigma_{j}(v_{r},\vartheta)}{d \Omega}.
\label{clrate} \end{equation}
These quantities also satisfy requirement (\ref{gcon}) and have the
discussed probabilistic interpretation. The collisional (classical,
center-of-mass) cross sections $d\sigma_{j}/d\Omega$ may obviously be
viewed as classical counterparts of scattering amplitudes appearing
in "diagonal" definitions (\ref{qmker}) and (\ref{qmrate}). There are
no classical analogs of coherences, so there is no classical argument
leading to coherence kernels and rates. Atomic coherences describe
the correlations between the internal quantum states of active atoms
and usually may be left out of the classical picture.

Kinetic equations (\ref{linbol}) or their semiclassical counterparts,
is then used as an augmentations of the equations of motion
describing the radiative effects occurring due to the coupling of the
active atoms with the incident (laser) radiation. Such an approach is
frequently used in the studies light-induced kinetic effects in gases
\cite{gpps,priv} and of other phenomena of the spectroscopic nature
in which atom-atom collisions are of importance (e.g. \cite{shap}).
However, computation of the collision kernels and rates either in
quantum-mechanical or in classical case is usually extremely
difficult, if at all possible. Calculations of the scattering
amplitudes (or equivalently of the cross sections) require the
knowledge of interatomic potentials. Such potentials are either not
known, or given only in the numerical form. Therefore, many authors
employ model kernels and rates which allow analytical computations.
The most popularly used models are strong collision one and the
Keilson-Storer model. There are also other possibilities connected
with the notion of collision operators. These models and concepts are
reviewed elsewhere, see for example \cite{rash,khab,kgon} and
references given therein.

\subsection{Comparison to our results}

We are now in position to compare our results with those above
presented which are known from literature. Although we focus our
attention on the Bloch-Boltzmann equations, i.e., on those for a
two-level atom, it seems that the generalizations to more complex
atomic structures are rather straightforward.

Adopting the operator basis (\ref{psd}) with the
hamiltonian (\ref{htla}) and leaving the inelastic collisions out of
the picture, which corresponds to putting
$T_{3}(\vv_{r} \leftarrow \vv_{r}') = %
T_{4}(\vv_{r} \leftarrow \vv_{r}') = 0$, we arrive at the BBE exactly
as in Eqs.(\ref{bbn}). Since we have an identification (\ref{roel})
we can conclude that our collision kernels $\ckk{ab}$ coincide with
$\wkk{ij}$, for $a=i$ and $b=j$, they are given by the same
formulas, Eqs.(\ref{ker1}) or (\ref{ker2}) and (\ref{qmker}),
respectively. Hence, the comments given after Eq.(\ref{qmrate}) apply
to population kernels $\wkk{ii}$ as well as to $\ckk{aa}$ for
$i=a=1,2$,

The situation is, however, much different with respect to our
collisional rates $\gamma^{\ast}_{ab}$ as compared to literature ones
$\gamma_{ij}$. First of all, we note that in our case the rates
$\gamma^{\ast}_{ab}$ are defined as integrals over kernels $\ckk{ab}$
and follow quite naturally from the general formalism. Hence,
$\gamma^{\ast}_{ab}$ are given as integrals over the products of
scattering amplitudes. The rates $\gamma_{ij}$ from literature are
given by differences of forward scattering amplitudes, see
Eq.(\ref{qmrate}). The derivation and the limiting procedures, as
presented by Rautian and Shalagin, which lead to the expression
(\ref{qmrate}) seem to lack mathematical rigor and their physical
basis seems to us to be not fully clear. Luckily, the application of
the optical theorem entails correct relation (\ref{gcon}) which, on
the other hand, is an automatic consequence of our formalism.

Secondly, as it follows from the very structure of our theory,
Bloch-Boltzmann equations for coherences, that is Eqs. (\ref{bbnc})
and (\ref{bbnd}) contain the coherence collisional rate, which is
in our case, of the form
\begin{equation}
   \Gamma_{12}(\vv) = \Gamma_{21}(\vv)
   = \half \leftk \gamma^{\ast}_{11}(\vv)
                + \gamma^{\ast}_{22}(\vv) \right].
\label{gcoh} \end{equation}
Such a quantity does not appear in the kinetic equation approach. The
presence of $\Gamma_{12}$ is connected with the employed method which
ensures the positivity of the atomic density operator. The kinetic
equation method which provides coherence rate $\gamma_{12}(\wek{v})$
cannot, therefore, guarantee that the computed density operator will
be positively defined, as it always should be.

Moreover, the rate $\Gamma_{12}(\wek{v})$ is specified only by
collision kernels $\ckk{11}$ and $\ckk{22}$ which may be called
"population" ones. $\Gamma_{12}(\wek{v})$ is, thus, independent of the
coherence kernel $\ckk{12}$ which enters Eqs.(\ref{bbn}) separately.
Only if inelastic collisions are taken into account, that is if the
scattering amplitudes $f_{3}(\vv_{r} \leftarrow \vv_{r}')$ and
$f_{4}(\vv_{r} \leftarrow \vv_{r}')$ are nonzero, the off-diagonal
rates $\gamma_{ab}^{\ast}$ may appear in Bloch-Boltzmann equations.
This is clearly seen by inspection of Eqs.(\ref{bbe}) which are the
most general ones. We are of the opinion that our
$\Gamma_{12}(\wek{v})$ should be associated with homogeneous
linewidth, usually connected with the so-called "phase-changing"
collisions. Having in view the relation between $\Gamma_{12}$ and
population collision kernels, we support the opinion of Rautian and
Shalagin (\cite{rash}, p.66), that any attempts to distinguish
"phase-changing collisions", "velocity-changing collisions" or any
other special types of collisions, are futile and physically
misleading. Thus, it also seems that attempts to split the
collisional rate as in Eq.(\ref{cohkin}) are arbitrary and not really
justified.

Furthermore, if we assume that the coherence kernel $\ckk{12}$ can be
neglected (putting aside the question of justification of such an
approximation) the rate $\Gamma_{12}(\wek{v})$ appears naturally in
our formalism. There is no need to invoke any additional arguments
that homogeneous linewidth should be present in Bloch-Boltzmann
equations, even when the kernel $\ckk{12}$ is absent. This conclusion
remains valid, also in the case we adopt semiclassical picture taking
classical expressions (\ref{clrate}) and (\ref{clker}) for
$\gamma^{\ast}_{aa}$ and $\ckk{aa}$.

Finally, we note that relation (\ref{gcon}) which is the requirement
of particle number conservation follows automatically from our
derivation, as it is reflected by expression (\ref{kgam}). This is
especially important in various approximate approaches to estimate
the collision kernels. In such approximations the optical theorem may
not hold, and thus (\ref{gcon}) may also not hold. The kinetic
equation method may not ensure the particle number conservation while
our approach clearly does, even if some approximation are made.

We conclude this work, by saying that we believe that the proper form
of the Bloch-Boltzmann equations is expressed by the set (\ref{bbn})
with collision kernels defined by formula (\ref{ker1}) or
(\ref{ker2}) and with collision rates connected to kernels by
relation (\ref{kgam}). The proposed proper form of BBE ensures that
the required positivity of the atomic density operator is preserved.
Moreover, we believe that the presented interpretation of the
structure and terms of BBE will appear useful in future applications
and in further research.

\begin{acknowledgments}
Partial support by Gda\'{n}sk University through grant
BW/5400-5-0158-1 is gratefully acknowledged.
\end{acknowledgments}

\end{document}